\documentclass[preprint]{article}
\usepackage{authblk}
\usepackage{graphicx}
\usepackage{dcolumn}
\usepackage{bm}
\usepackage{amsmath}
\usepackage{amssymb}
\usepackage{multirow}
\usepackage{caption}
\usepackage{subcaption}
\usepackage{epstopdf}
\usepackage{hyperref}
\usepackage{cite}
\begin{document}
\title{\bf Late-time cosmic dynamics in $f(R,L_{m})$ gravity with recent observations}
\author[]{Amit Samaddar\thanks{samaddaramit4@gmail.com}}
\author[]{S. Surendra Singh\thanks{ssuren.mu@gmail.com}}
\affil[]{Department of Mathematics, National Institute of Technology Manipur,\\ Imphal-795004, India.}
\maketitle

\begin{abstract}
In this work, we investigate the late-time cosmic dynamics in the framework of non-linear $f(R, L_m)$ gravity, adopting the functional form $f(R,L_m)=\frac{R}{2}+L_m^2$. To explore the dark energy behavior, we assume an oscillatory parametric equation of state, $\omega(z) = \omega_0 + b \sin[\log(1+z)]$, which allows smooth deviations from the cosmological constant. Using a joint MCMC analysis with the latest Hubble 31 chronometer data, DESI DR2 BAO measurements, and Type Ia supernova samples (Pantheon+, DES-SN5Y and Union 3), we obtain well-constrained parameters around $H_0 \simeq 67.2~\text{km s}^{-1}\text{Mpc}^{-1}$ and $\omega_0\approx-0.5$, consistent with Planck 2018 and other current observations. The model exhibits a clear transition from deceleration to acceleration with $z_{\rm tr} \sim 0.7$--$0.8$, satisfies the NEC and DEC while violating the SEC and yields present EoS values close to $-1$, reproducing $\Lambda$CDM behavior at late times. The derived Universe ages ($t_0 \approx 13.3~\text{Gyr}$) agree well with CMB and stellar constraints, confirming that the proposed oscillatory $f(R, L_m)$ model provides an observationally consistent and dynamically viable alternative to $\Lambda$CDM cosmology.
\end{abstract}


\section{Introduction}\label{sec1}
\hspace{0.5cm} Modern cosmology aims to understand the origin, large-scale structure, and long-term evolution of the Universe. In recent decades, increasingly precise astronomical observations have dramatically reshaped our comprehension of cosmic dynamics, revealing that the Universe is not only expanding but undergoing accelerated expansion. This phenomenon was first identified through high-redshift type Ia supernovae (SNe Ia) observations and subsequently corroborated by observations of Cosmic Microwave Background (CMB) anisotropies, Baryon Acoustic Oscillations (BAO) and large-scale structure (LSS) surveys \cite{BP1,BP2,BP3,BP4,BP5}. Collectively, these datasets indicate a spatially flat Universe dominated by an unknown dark energy component responsible for the late-time acceleration. The standard $\Lambda$CDM model describes this evolution by incorporating a cosmological constant ($\Lambda$) as dark energy and cold dark matter (CDM) to account for structure formation. While $\Lambda$CDM matches observational results remarkably well, it suffers from conceptual challenges, including the fine-tuning of $\Lambda$, the cosmic coincidence problem, and the absence of a fundamental explanation for dark energy. Additional issues, such as the initial singularity problem and the observed tension in the Hubble constant $H_0$ as determined from both early- and late-universe probes suggest that a deeper understanding of the underlying gravitational framework may be required \cite{SM2001}.

General Relativity (GR), developed by Einstein in 1915, remains the foundation of modern gravitational theory and has successfully explained a wide range of astrophysical and cosmological phenomena \cite{A1917}. Its predictions, including the perihelion precession of Mercury, the deflection of light by gravity, and the recent detection of gravitational waves by the LIGO and Virgo collaborations, have been confirmed with high precision. On cosmological scales, however, GR requires the introduction of unknown components, such as dark matter and dark energy, to account for the observed accelerated expansion and the structure of the Universe. These limitations, along with conceptual issues like singularities and the cosmological constant problem, have motivated the development of alternative theories of gravity that could provide a deeper understanding of cosmic dynamics. To address these limitations, numerous modified gravity theories have been proposed, extending or generalizing Einstein's formulation to incorporate higher-order curvature terms, scalar fields, or direct couplings between matter and geometry. These theories aim to provide a purely geometric origin for cosmic acceleration without invoking a separate dark energy component. Among them, $f(R)$ gravity stands out as one of the most natural extensions of GR, where the Einstein--Hilbert Lagrangian $R$ is generalized to a function $f(R)$ of the Ricci scalar. This modification introduces additional degrees of freedom that can effectively drive both the inflationary and late-time accelerating phases of the Universe, offering a unified cosmological description \cite{Kerner1982,HA1970}.

A further generalization arises in the form of $f(R,L_m)$ gravity, where $L_m$ represents the matter Lagrangian density. The inclusion of a non-minimal coupling between matter and geometry introduces new dynamical effects, leading to non-geodesic motion of massive particles and a richer phenomenology at cosmological scales. This coupling can mimic the behavior of dark energy through the interaction between matter and curvature, potentially explaining the accelerated expansion in a purely geometrical framework. Such models have been extensively studied for their cosmological implications, energy conditions and compatibility with observational data, making them a promising avenue toward understanding the nature of cosmic acceleration and the interplay between matter and geometry \cite{VF2004,Bero06,Wang2012}. In recent years, several studies have explored the cosmological implications of the $f(R,L_m)$ framework, highlighting its versatility in addressing various aspects of cosmic evolution. In \cite{LV2022}, the authors examined the general cosmological dynamics within this theory, while \cite{LV2023} investigated the mechanism of baryogenesis in the same framework. A constrained cosmological model consistent with observational data was proposed in \cite{singh2023}, whereas \cite{Maurya2023} analyzed accelerating scenarios in massive cosmology under $f(R,L_m)$ gravity. More recently, \cite{Devi2024} provided parameter constraints for an accelerating universe within this formulation. These studies collectively demonstrate the rich phenomenology and observational relevance of $f(R,L_m)$ gravity in explaining late-time cosmic acceleration without the need for an explicit dark energy component.

The equation of state (EoS), defined as $\omega(z) = \frac{p}{\rho}$, is a fundamental quantity in cosmology that describes the dynamical evolution of the accelerating Universe. Since the physical origin of dark energy remains unknown, a widely used approach is to model $\omega(z)$ through phenomenological parameterizations. Such forms offer a flexible, model-independent framework for confronting theoretical predictions with current observational data \cite{Tpad03}. The simplest scenario assumes a constant EoS, $\omega(z) = \omega_0$, which encompasses the cosmological constant ($\Lambda$CDM, $\omega_0 = -1$) and its straightforward generalization, the $\omega$CDM model. To account for redshift-dependent behavior, the Chevallier–Polarski–Linder (CPL) parametrization \cite{Chevallier2001,Linder2003}, $\omega(z) = \omega_0 + \omega_a \frac{z}{1+z}$, provides a smooth interpolation between the present epoch and high-redshift limits. Alternative representations include the linear form $\omega(z)=\omega_0+\omega_1 z$ \cite{Huterer2001}, the logarithmic model $\omega(z)=\omega_0+\omega_1 \ln(1+z)$ \cite{Efstathiou1999}, and bounded expressions like the Barboza–Alcaniz model \cite{Barboza2008}, $\omega(z)=\omega_0+\omega_1 \frac{z(1+z)}{1+z^2}$, and the Jassal–Bagla–Padmanabhan (JBP) form \cite{Jassal2005}, $\omega(z)=\omega_0+\omega_1 \frac{z}{(1+z)^2}$. Other parametrizations, such as Wetterich’s $\omega(z)=\frac{\omega_0}{1+b\ln(1+z)}$ \cite{Wetterich2004} and Gong–Zhang’s $\omega(z)=\frac{\omega_0}{1+z}$ \cite{Gong2005}, capture tracking behaviors or remain well-defined across all redshifts. Building on these ideas, we adopt a generalized ansatz \cite{Pan18}, $\omega(z) = \omega_0 + b \left[1 - \cos\left(\ln(1+z)\right)\right]$, which allows smooth, bounded variations simulating transient departures from $\Lambda$CDM. In this study, we investigate a nonlinear gravity model, $f(R, L_m) = \frac{R}{2} + L_m^2$, combined with an oscillatory EoS, $\omega(z) = \omega_0 + b \sin[\ln(1+z)]$, and constrain its parameters using a Markov Chain Monte Carlo (MCMC) analysis with the latest Hubble 31 cosmic chronometer measurements, DESI DR2 BAO data, and Type Ia supernova compilations including Pantheon+, DES-SN5Y, and Union 3, providing a robust observational assessment of the proposed framework.

The structure of this paper is organized as follows. In section \ref{sec2}, we present the theoretical framework of $f(R, L_m)$ gravity, focusing on the geometry--matter coupling and the corresponding field equations. Section \ref{sec3} is devoted to the formulation of the specific $f(R, L_m)$ model adopted in this work, along with the assumed EoS form. In this section, we also solve the field equations and derive the analytical expression for the Hubble parameter $H(z)$. In Section \ref{sec4}, we utilize the latest observational datasets, including Hubble chronometers, BAO, and Type Ia supernovae, to perform parameter estimation and determine the best-fit values of the model parameters. Section \ref{sec5} focuses on the investigation of essential cosmological parameters and their role in shaping the Universe’s expansion history. In section \ref{sec6}, we conclude by summarizing our results and highlighting their relevance within the framework of modified gravity and the study of late-time cosmic acceleration.
\section{Geometry–matter coupling and field equations in $f(R,L_{m})$ gravity}\label{sec2}
\hspace{0.5cm} In this section, we provide a concise summary of the modified theory of gravity formulated within the $f(R,L_{m})$ framework. The corresponding action for this model can be written as \cite{Harko2010}
\begin{equation}\label{1}
S=\int f(R,L_{m})\sqrt{g}d^{4}x,
\end{equation}
Here, $L_{m}$ denotes the matter Lagrangian density, while $R$ represents the Ricci scalar. The Ricci scalar $R$ is constructed from the Ricci tensor $R_{\mu\nu}$ and the metric tensor $g^{\mu\nu}$ and is given by  
\begin{equation}\label{2}
R = g^{\mu\nu} R_{\mu\nu}.
\end{equation}
The Ricci tensor $R_{\mu\nu}$ is formulated through the connection coefficients as  
\begin{equation}\label{3}
R_{\mu\nu}=\partial_{\lambda} \Gamma^{\lambda}_{\mu\nu}-\partial_{\nu} \Gamma^{\lambda}_{\mu\lambda}+\Gamma^{\lambda}_{\mu\nu} \Gamma^{\sigma}_{\lambda\sigma}-\Gamma^{\sigma}_{\mu\lambda} \Gamma^{\lambda}_{\nu\sigma}.
\end{equation}

Here, $\Gamma^{\alpha}_{\beta\gamma}$ denotes the Christoffel symbols (Levi-Civita connection), defined as  
\begin{equation}\label{4}
\Gamma^{\alpha}_{\beta\gamma} = \frac{1}{2} g^{\alpha\lambda}\left(\frac{\partial g_{\lambda\beta}}{\partial x^{\gamma}}+\frac{\partial g_{\lambda\gamma}}{\partial x^{\beta}}-\frac{\partial g_{\beta\gamma}}{\partial x^{\lambda}}\right).
\end{equation}
A metric variation of the action (\ref{1}) with respect to $g_{\mu\nu}$ yields the corresponding field equations for the $f(R, L_{m})$ theory.
\begin{equation}\label{5}
f_{R} R_{\mu\nu}+ \left(g_{\mu\nu}\Box-\nabla_{\mu} \nabla_{\nu}\right)f_{R}-\frac{1}{2}\left(f-f_{L_{m}}L_{m} \right)g_{\mu\nu}=\frac{1}{2} f_{L_{m}} T_{\mu\nu},
\end{equation}
where $\Box = g^{\mu\nu} \nabla_{\mu} \nabla_{\nu}$, $f_{R} \equiv \dfrac{\partial f}{\partial R}$ and $f_{L_{m}} \equiv \dfrac{\partial f}{\partial L_{m}}$. The quantity $T_{\mu\nu}$ denotes the energy--momentum tensor of the matter content. For a general matter Lagrangian $L_m$, the energy--momentum tensor is defined as
\begin{equation}\label{6}
T_{\mu\nu} = -\frac{2}{\sqrt{-g}} \frac{\delta \left( \sqrt{-g}\, L_m \right)}{\delta g^{\mu\nu}}.
\end{equation}
The field equations allow us to formulate a connection between the Ricci scalar $R$, the energy--momentum tensor trace $T$ and the matter Lagrangian density $L_m$, expressed as
\begin{equation}\label{7}
R f_{R}+2\left(f_{L_{m}}L_m-f\right) + 3 \, \Box f_{R}=\frac{1}{2} f_{L_{m}}T,
\end{equation}
In the above, the d'Alembert operator $\Box$ acting on an arbitrary scalar function $I$ is defined by
\begin{equation}\label{8}
\Box I = \frac{1}{\sqrt{-g}} \, \partial_\mu \Big( \sqrt{-g} \, g^{\mu\nu} \, \partial_\nu I \Big),
\end{equation}
which provides the covariant generalization of the Laplacian in curved spacetime.

By inserting the covariant derivative relation for the energy--momentum tensor into equation (\ref{5}), we find
\begin{equation}\label{9}
\nabla_\mu T^{\mu\nu}=2 \, \nabla_\mu \ln\left(f_{L_{m}}\right) \frac{\partial L_m}{\partial g_{\mu\nu}}.
\end{equation}
This result indicates that the energy--momentum tensor is generally not conserved in $f(R, L_m)$ gravity, due to the explicit coupling between the matter Lagrangian density and the spacetime geometry. The non-vanishing divergence reflects a transfer of energy and momentum between the matter and gravitational sectors, illustrating a key difference from standard general relativity.

Observations of the Universe indicate that, on large scales, the current energy density closely matches the value predicted by a spatially flat cosmological model. For the purpose of studying the Universe’s evolution, we adopt a spatially flat FLRW metric, formulated as
\begin{equation}\label{10}
ds^2=-dt^2+a^2(t)\,(dx^2+dy^2+dz^2),
\end{equation}
where $a(t)$ denotes the scale factor describing the expansion of the Universe.

When equation (\ref{3}) is applied to the spatially flat FLRW metric, the Ricci tensor’s nonzero quantities take the form
\begin{equation}\label{11}
R_{00} = 3 \frac{\ddot a}{a}, \quad R_{11} = R_{22} = R_{33} = \frac{\ddot a}{a} + 2 \left( \frac{\dot a}{a} \right)^2,
\end{equation}
where $a(t)$ is the scale factor.

From these components, the Ricci scalar for the metric in equation (\ref{9}) takes the form
\begin{equation}\label{12}
R = 6 \left[ \frac{\ddot a}{a} + \left( \frac{\dot a}{a} \right)^2 \right] = 6 \left( 2H^2 + \dot H \right),
\end{equation}
with $H =\frac{\dot{a}}{a}$ representing the Hubble expansion rate. This formulation explicitly relates the spacetime curvature to the expansion dynamics of the Universe.

The energy--momentum tensor describing a perfect cosmic fluid, characterized by the energy density $\rho$ and isotropic pressure $p$ (neglecting viscosity), is given by
\begin{equation}\label{13}
T_{\mu\nu}=(\rho+p)\,u_\mu u_\nu+p\,g_{\mu\nu},
\end{equation}
where $u_\mu = (1,0,0,0)$ represents the four-velocity of the comoving cosmic fluid in the FLRW spacetime.

The equations dictating the Universe’s dynamics under $f(R, L_m)$ gravity take the form
\begin{equation}\label{14}
3 H^2 f_R + \frac{1}{2} \left( f - R f_R - f_{L_m} L_m \right) + 3 H \dot{f}_R = \frac{1}{2} f_{L_m} \, \rho,
\end{equation}
and
\begin{equation}\label{15}
\dot{H} f_R + 3 H^2 f_R - \ddot{f}_R - 3 H \dot{f}_R + \frac{1}{2} \left( f_{L_m} L_m - f \right) = \frac{1}{2} f_{L_m} \, p,
\end{equation}
\section{Analysis of a non-linear $f(R, L_m)$ model and its cosmological implications}\label{sec3}
\hspace{0.5cm} To study the effects of a non-minimal interaction between spacetime geometry and matter, we examine a particular nonlinear form of the gravitational Lagrangian \cite{Nmyr2023} expressed as 
\begin{equation}\label{16}
f(R, L_m)=\frac{R}{2}+L_{m}^{2},
\end{equation}
This choice constitutes one of the most straightforward extensions of general relativity, featuring a quadratic dependence on the matter Lagrangian alongside a linear term in the Ricci scalar $R$. The linear curvature term guarantees that the theory recovers standard GR in the weak-field regime, whereas the $L_m^2$ component introduces nonlinear modifications capable of capturing the effects of matter--geometry coupling beyond the conventional Einstein--Hilbert framework. The rationale for considering this model is grounded in both physical and theoretical motivations. The quadratic dependence on $L_m$ naturally incorporates self-interaction effects in matter, which can emulate dark energy-like behavior without invoking additional exotic fields. Additionally, such nonlinear matter--geometry couplings have been demonstrated to generate late-time cosmic acceleration and lead to deviations from the usual conservation of the energy--momentum tensor, offering potential explanations for observed cosmological phenomena. Moreover, the model is mathematically manageable, allowing for analytic solutions of the field equations and facilitating direct comparison with observations via the evolution of the Hubble parameter $H(z)$.

By choosing the matter Lagrangian density as $L_m=\rho$ \cite{Smyr2024} and substituting it into the selected nonlinear model, the modified field equations can be explicitly solved. In this scenario, the derivatives of the function $f(R, L_m)$ are given by $f_R = \frac{\partial f}{\partial R} = \frac{1}{2}$ and $f_{L_m} = \frac{\partial f}{\partial L_m} = 2 L_m = 2 \rho$. These expressions provide the necessary ingredients to analyze the cosmic dynamics within this specific framework. The resulting equations governing the cosmic evolution can be written as:
\begin{equation}\label{17}
\rho(z)=H(z), \quad 2\dot{H}+3H^{2}=\rho^{2}-2\rho p.
\end{equation}
From the above relation, one can derive the expression for the equation of state (EoS) parameter $\omega$ as follows:
\begin{equation}\label{18}
\omega=\frac{p}{\rho}=\frac{1}{2}-\frac{2\dot{H}+3H^{2}}{2H^{2}},
\end{equation}

To investigate the dynamics of the cosmic fluid within our nonlinear model $f(R, L_m) = \frac{R}{2} + L_m^2$, we adopt a redshift-dependent EoS parameter of the form
\begin{equation}\label{19}
\omega(z) = \omega_0 + b \, \sin \big( \log(1+z) \big),
\end{equation}
where $\omega_0$ denotes the present-day value of the EoS and $b$ controls the amplitude of the oscillatory deviation. This parametrization introduces a gentle, slowly varying oscillation around $\omega_0$, allowing the EoS to evolve dynamically with redshift while remaining regular at early times. Oscillatory EoS forms have been explored in the literature to capture possible deviations from a constant dark energy scenario and to model transitions between different phases of cosmic acceleration \cite{Lazkoz2010,Pace2012,Demianski2020}. In our framework, the combination of this oscillatory EoS with the nonlinear matter coupling offers a versatile approach to study the effects of matter--geometry interaction on the cosmic expansion history, providing a consistent method to determine the evolution of the Hubble parameter $H(z)$.

By substituting the assumed EoS form from equation (\ref{18}) into equation (\ref{16}) and using the relation $\dot{H} = -H(z)(1+z)\,\frac{dH}{dz}$, we obtain a differential equation for the Hubble parameter. Solving this equation provides an explicit redshift-dependent expression for $H(z)$, which describes the cosmic evolution within the chosen $f(R, L_m)$ framework.
\begin{equation}\label{20}
H(z)=H_{0}(1+z)^{1+\omega_{0}}e^{-b\cos(\log(1+z))}\;,
\end{equation}
where $H_{0}$ is the present value of the Hubble parameter.
\section{Parameter constraints from cosmological observables}\label{sec4}
\hspace{0.5cm} The objective of this section is to employ the most recent and comprehensive observational datasets to constrain the free parameters of the proposed $f(R, L_m)$ model. To achieve this, we utilize the Markov Chain Monte Carlo (MCMC) technique within a Bayesian statistical framework, allowing an efficient exploration of the multidimensional parameter space. The analysis is implemented using the Python package \texttt{emcee} \cite{Mackey13}, which is a widely used and robust sampler for obtaining posterior probability distributions of cosmological parameters. In this framework, the likelihood function is assumed to follow a Gaussian form,
\begin{equation}\label{21}
\mathcal{L} \propto \exp\left(-\frac{\chi^{2}}{2}\right),
\end{equation}
where $\chi^{2}$ measures the deviation between the model predictions and the corresponding observational data points. For this study, the free parameters $(H_0, \omega_0, b)$ are allowed to vary uniformly within the intervals
\begin{equation}\label{22}
50 < H_0 < 100, \quad -1 < \omega_0 < 1, \quad 0 < b < 2.
\end{equation}
These prior ranges are consistent with current cosmological constraints while remaining broad enough to ensure an unbiased parameter search. The MCMC chains are initialized with random values drawn from the prior distributions and evolved using a sufficiently large number of walkers and iterations to guarantee statistical convergence. The stability and efficiency of the sampling are verified through diagnostics such as the autocorrelation time.

For parameter estimation, we perform a joint analysis using multiple independent datasets that probe the late-time expansion history of the Universe. Specifically, we incorporate the \textit{Hubble 31} cosmic chronometer measurements, the \textit{DESI DR2} baryon acoustic oscillation (BAO) data, the \textit{Pantheon+} and \textit{DES-SN5Y} Type Ia supernova compilations, along with the newly released \textit{Union 3} dataset. The combined likelihood function from these datasets provides stringent constraints on the model parameters, enabling a consistent evaluation of the $f(R, L_m)$ framework against current cosmological observations. The resulting posterior distributions yield both the best-fit values and the corresponding credible intervals for each free parameter.
\subsection{Cosmic chronometer (CC31) dataset analysis}\label{sec4.1}
\hspace{0.5cm} To evaluate the observational consistency of our cosmological framework, we employ the Cosmic Chronometer (CC31) dataset, which provides 31 independent determinations of the Hubble parameter $H(z)$. These values are derived using the differential age (DA) approach applied to passively evolving galaxies spanning the redshift interval $0.07 \leq z \leq 1.965$. This technique offers a largely model-independent estimate of the expansion rate, as it relies on the relative aging of galaxies with redshift rather than presuppositions regarding the cosmological model. Following the methodology outlined in \cite{M12,Sam24,ASamaddar25}, the Hubble parameter can be expressed as  
\begin{equation}\label{23}
H(z) = -\frac{1}{1+z}\frac{dz}{dt},
\end{equation}
which establishes a direct connection between the observable change in redshift and the rate of cosmic time evolution.

The model parameters $\theta=(H_{0},\omega_{0},b)$ are constrained by minimizing a statistical measure that quantifies the difference between the theoretical Hubble expansion and the observed CC data. The chi-square for this dataset is expressed as
\begin{equation}\label{24}
\chi^{2}_{\mathrm{CC}}(\theta) = \sum_{i=1}^{31} \frac{\left[ H_{\mathrm{th}}(\theta, z_i) - H_{\mathrm{obs}}(z_i) \right]^2}{\sigma_{H}^{2}(z_i)},
\end{equation}
where $H_{\text{th}}(\theta, z_i)$ represents the theoretical value of the Hubble parameter predicted by the model for a given parameter set $\theta$, $H_{\text{obs}}(z_i)$ denotes the corresponding observational measurements, and $\sigma_H(z_i)$ is the standard deviation associated with each data point.
\subsection{Incorporation of DESI DR2 BAO Dataset}\label{sec4.2}
\hspace{0.5cm} To improve the precision of our cosmological parameter estimation, we incorporate the newly released Baryon Acoustic Oscillation (BAO) results from the DESI Data Release 2 (DR2). By analyzing the clustering of over 14 million galaxies and quasars across a vast redshift range, DESI DR2 delivers some of the most detailed and reliable distance measurements to date, making it a key resource for contemporary BAO studies \cite{DESI25}. The DESI DR2 catalogue includes multiple tracer populations, namely the Bright Galaxy Sample (BGS), Luminous Red Galaxies (LRG1, LRG2, LRG3+ELG1), Emission Line Galaxies (ELG2), Quasars (QSO), and Lyman-$\alpha$ forests (Ly$\alpha$). These tracers jointly inform the BAO analysis by providing constraints on several key distance ratios, including $D_M/r_d$, $D_H/r_d$, $D_V/r_d$ and $D_M/D_H$. The parameter $r_d$ represents the sound horizon at the drag epoch, taking the value $r_d = 147.09 \pm 0.20\,\mathrm{Mpc}$ under the standard $\Lambda$CDM model. From a theoretical perspective, the corresponding distance indicators can be expressed as functions of the Hubble parameter $H(z)$ of our model:
\begin{equation}\label{25}
D_{M}(z) = c \int_{0}^{z} \frac{dz'}{H(z')},
\end{equation}
\begin{equation}\label{26}
D_{H}(z) = \frac{c}{H(z)},
\end{equation}
and the volume-averaged distance is given by
\begin{equation}\label{27}
D_{V}(z) = \left[ z D_{M}^{2}(z) D_{H}(z) \right]^{1/3}.
\end{equation}

The model parameters $\theta = (H_0, \omega_0, b)$ are estimated by comparing the theoretical predictions with the DESI DR2 BAO data through the chi-squared function,
\begin{equation}\label{28}
\chi^{2}_{\mathrm{BAO}}(\theta) = \sum_{i} \frac{\left[ D^{\mathrm{th}}_{i}(\theta, z_i) - D^{\mathrm{obs}}_{i}(z_i) \right]^2}{\sigma_{i}^{2}},
\end{equation}
where $D^{\mathrm{obs}}_{i}(z_i)$ and $D^{\mathrm{th}}_{i}(\theta, z_i)$ correspond to the observed and theoretical BAO measurements, respectively, and $\sigma_i$ represents their measurement errors.
\subsection{Incorporation of Type Ia Supernova datasets: Pantheon+, DES-SN5Y and Union 3}\label{sec4.3}
\hspace{0.5cm} For more precise cosmological parameter estimation, we incorporate three extensive Type Ia supernova (SNIa) datasets, —Pantheon+, DES-SN5Y and Union 3, which provide standard candle measurements that probe the Universe’s late-time expansion history. The Pantheon+ dataset \cite{Brout22} is among the most precise and homogeneous compilations of Type Ia supernovae to date, comprising $1701$ light curves from $1550$ supernovae spanning the redshift range $0.01 < z < 2.26$. In our analysis, we exclude events with $z < 0.01$ to minimize systematic uncertainties arising from local peculiar velocities. The DES-SN5Y dataset \cite{DES25} comprises 1829 photometric light curves collected over the complete five-year Dark Energy Survey Supernova program. It includes 1635 supernovae discovered by DES and 194 external low-redshift events from the CfA and CSP compilations. This dataset offers independent, high-precision photometric measurements that complement the Pantheon+ sample. The Union 3 compilation \cite{Rubin25} represents the most recent release from the Supernova Cosmology Project, comprising 2087 Type Ia supernovae drawn from 24 different surveys over the redshift range $0.01 < z < 2.26$. The dataset has been uniformly calibrated using the SALT3 light-curve model and processed with the UNITY1.5 Bayesian framework, yielding 22 high-quality binned distance modulus measurements.
For all SNIa datasets, the theoretical distance modulus is expressed as
\begin{equation}\label{29}
\mu_{\mathrm{th}}(z_i; \theta) = 5 \log_{10}\left[\frac{D_{L}(z_i; \theta)}{\mathrm{Mpc}}\right] + 25,
\end{equation}
where $\theta = (H_0, \omega_0, b)$ represents the set of free model parameters, and the luminosity distance $D_L(z; \theta)$ is given by
\begin{equation}\label{30}
D_{L}(z; \theta) = c (1 + z) \int_{0}^{z} \frac{dz'}{H(z'; \theta)}.
\end{equation}
The corresponding chi-squared estimator used for parameter fitting is
\begin{equation}\label{31}
\chi^{2}_{\mathrm{SN}}(\theta) = [\mu{\mathrm{th}}(\theta) - \mu_{\mathrm{obs}}]^{T} C_{\mathrm{SN}}^{-1} [\mu_{\mathrm{th}}(\theta) - \mu_{\mathrm{obs}}],
\end{equation}
where $C_{\mathrm{SN}}$ denotes the total covariance matrix accounting for both statistical and systematic uncertainties.

To improve the precision of the model parameters $(H_0, \omega_0, b)$, we carry out a joint chi-squared analysis that combines several observational datasets. The total chi-squared is defined as
\begin{equation}\label{32}
\chi^{2}_{\mathrm{tot}} = \chi^{2}_{\mathrm{CC}} + \chi^{2}_{\mathrm{BAO}} + \chi^{2}_{\mathrm{SN}},
\end{equation}
with $\chi^{2}_{\mathrm{SN}}$ representing the contributions from the Pantheon+, DES-SN5Y and Union 3 supernova compilations. Using this framework, we perform three separate joint analyses corresponding to the following combinations of datasets: CC + BAO + Pantheon+, CC + BAO + DES-SN5Y and CC + BAO + Union 3.

\begin{figure}[htbp]
    \centering
    \begin{subfigure}{0.45\textwidth}
        \includegraphics[width=\linewidth]{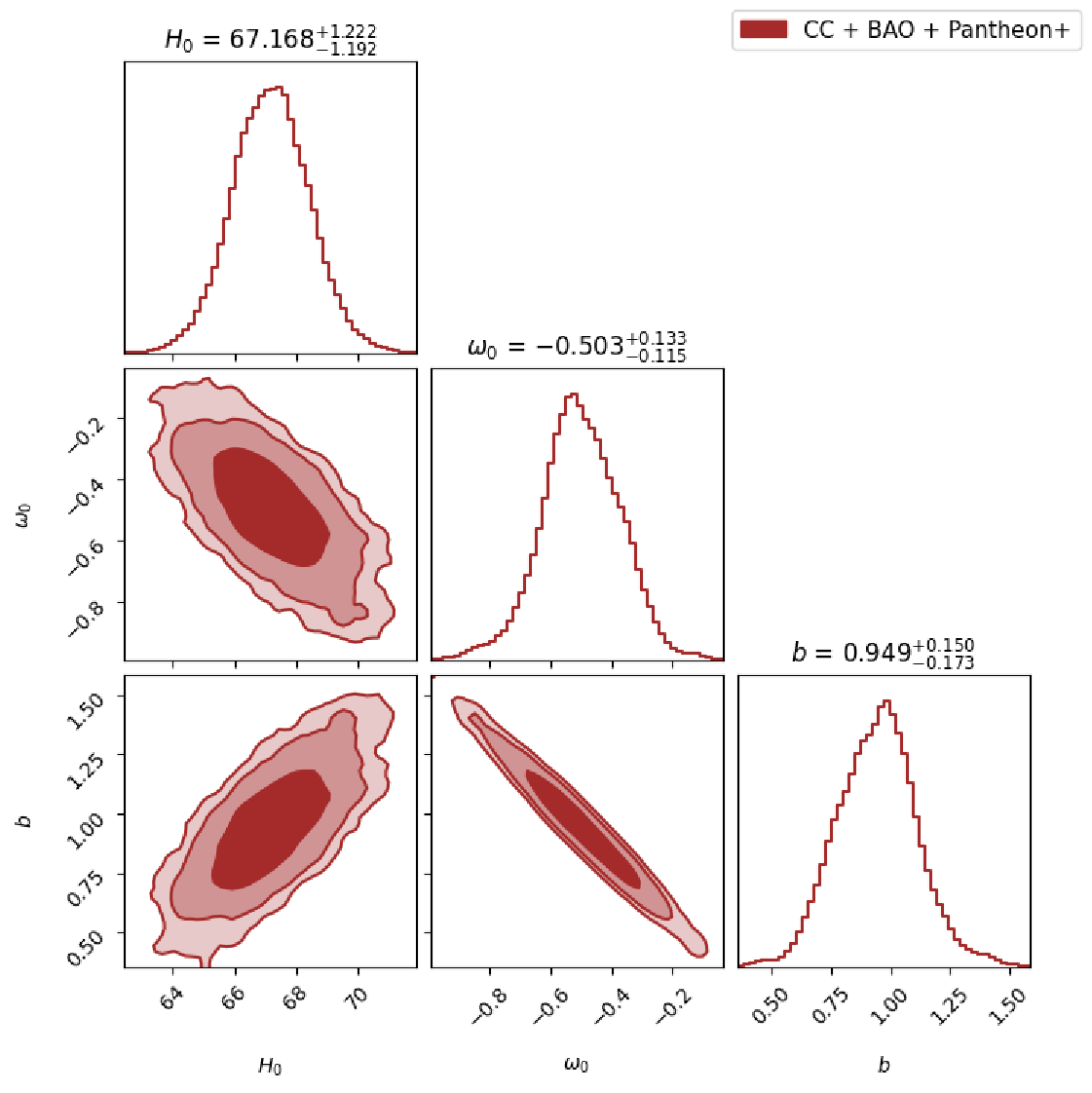} 
        \caption{CC + BAO + Pantheon+}
        \label{fig:subfig1}
    \end{subfigure}
    \hfill
    \begin{subfigure}{0.45\textwidth}
        \includegraphics[width=\linewidth]{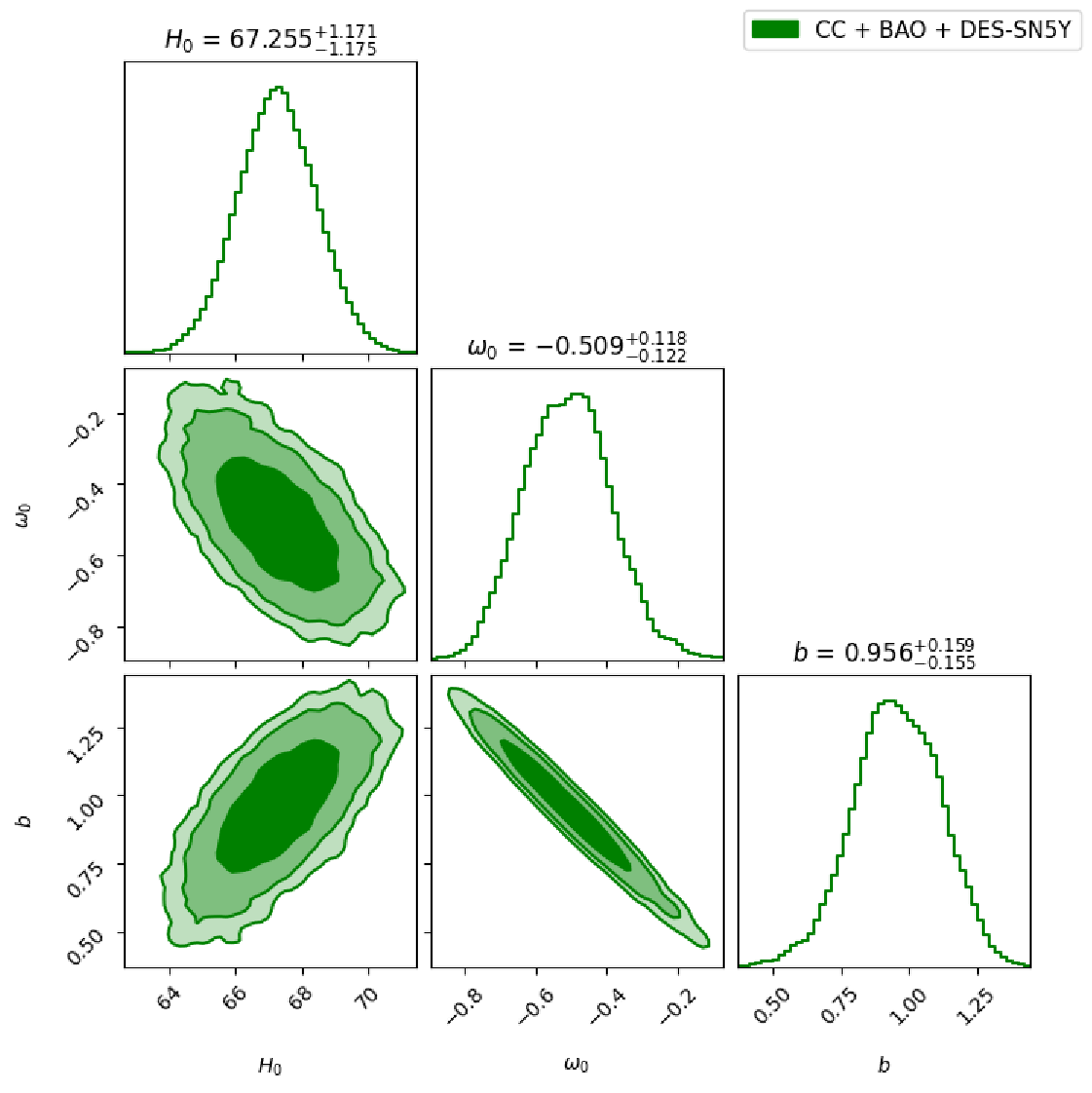} 
        \caption{CC + BAO + DES-SN5Y}
        \label{fig:subfig2}
    \end{subfigure}

    \begin{subfigure}{0.45\textwidth}
        \includegraphics[width=\linewidth]{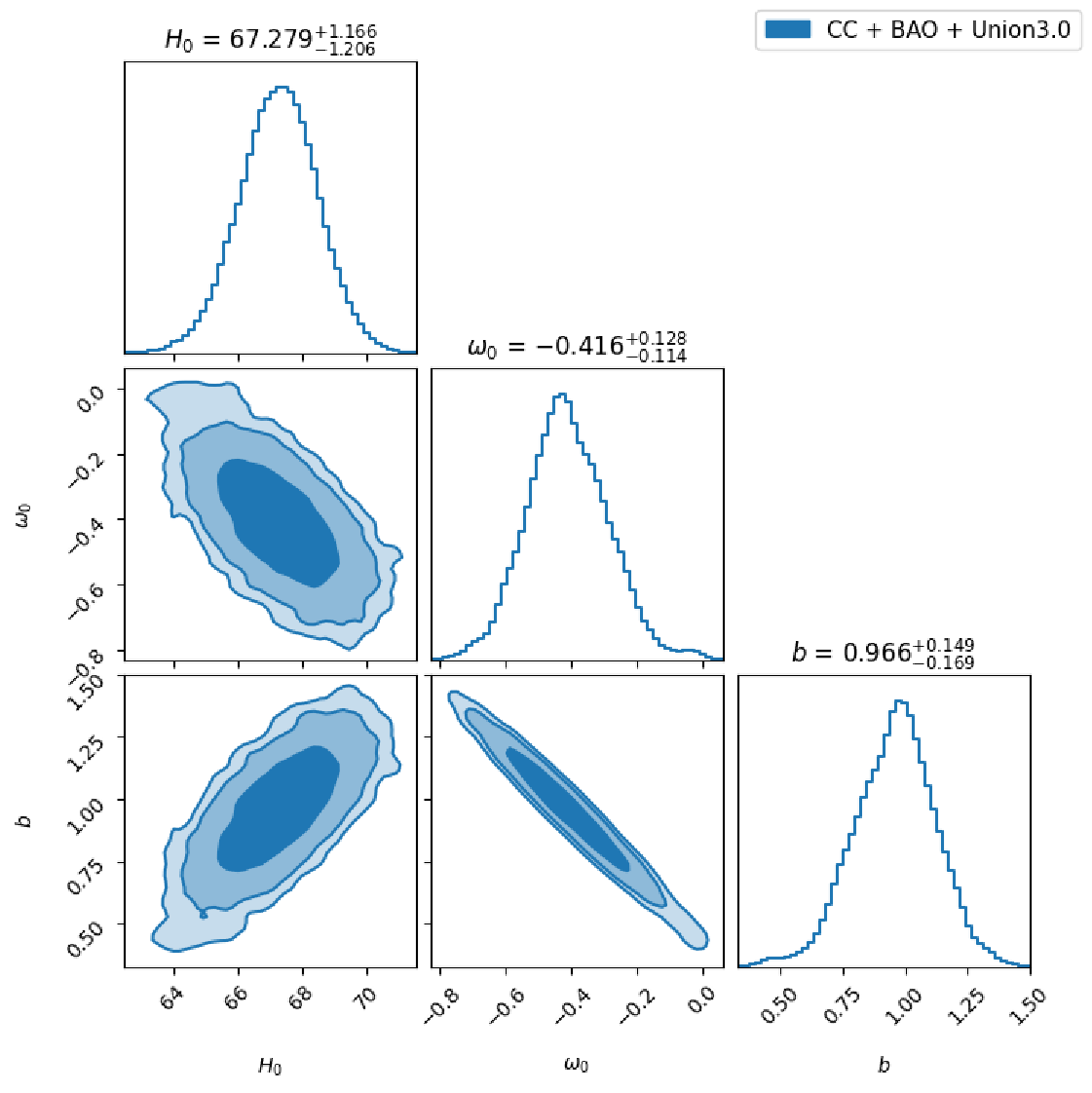} 
        \caption{CC + BAO + Union 3}
        \label{fig:subfig3}
    \end{subfigure}

    \caption{Marginalized one- and two-dimensional posterior distributions of the model parameters $(H_0, \omega_0, b)$ derived from the joint MCMC analysis using (a) CC+BAO+Pantheon+, (b) CC+BAO+DES-SN5Y and (c) CC+BAO+Union 3 datasets. The inner, middle and outer contours represent the $1\sigma$, $2\sigma$, and $3\sigma$ confidence intervals, respectively, illustrating the consistency and complementarity among the combined cosmological probes.}
    \label{fig:f1}
\end{figure}

Each joint dataset combination is analyzed within the same Bayesian framework described earlier, ensuring a consistent treatment of uncertainties. The resulting one-dimensional marginalized posterior distributions and two-dimensional confidence contours for the parameters $(H_0, \omega_0, b)$ are shown in Figure \ref{fig:f1}, displaying the $1\sigma$, $2\sigma$ and $3\sigma$ confidence regions corresponding to each dataset combination.
\begin{table}[h!]
\centering
\caption{Best-fit values of the model parameters $(H_0, \omega_0, b)$ obtained from the joint MCMC analysis.}
\begin{tabular}{||p{3.7cm}|p{3.1cm}|p{1.9cm}|p{1.5cm}|p{1cm}|p{1cm}|p{1cm}||}
\hline\hline
\hspace{0.9cm} Datasets & $H_0$(km s$^{-1}$ Mpc$^{-1}$) & \hspace{0.6cm} $\omega_0$ & \hspace{0.6cm}$b$ & \hspace{0.2cm}$\chi^{2}_{\text{min}}$ & $\Delta$AIC & $\Delta$BIC\\
\hline\hline
CC+BAO+Pantheon+ & \hspace{0.6cm}$67.168^{+1.222}_{-1.192}$ & $-0.503^{+0.133}_{-0.115}$ & $0.949^{+0.150}_{-0.173}$ & $796.73$ & $1.26$ & $11.21$\\[1.3pt]
\hline
CC+BAO+DES-SN5Y & \hspace{0.6cm}$67.255^{+1.171}_{-1.175}$ & $-0.509^{+0.118}_{-0.122}$ & $0.956^{+0.159}_{-0.155}$ & $1755.47$ & $1.21$ & $11.54$\\[1.3pt]
\hline
CC+BAO+Union 3 & \hspace{0.6cm}$67.279^{+1.166}_{-1.206}$ & $-0.416^{+0.128}_{-0.114}$ & $0.966^{+0.149}_{-0.169}$ & $48.82$ & $1.30$ & $11.50$ \\ 
\hline\hline
\end{tabular}
\label{Tab:T1}
\end{table}
\subsection{Discussion of results}\label{sec4.4}
\hspace{0.5cm} The best-fit parameters for our $f(R, L_m) = \frac{R}{2} + L_m^2$ model, combined with the oscillatory equation of state $\omega(z) = \omega_0 + b \sin[\log(1+z)]$, exhibit strong agreement with current cosmological observations. Across all dataset combinations, the resulting Hubble constant, $H_0 \simeq 67.2 \pm 1.2 \mathrm{km s^{-1} Mpc^{-1}}$, is in excellent agreement with the Planck 2018 CMB measurement ($H_0 = 67.4 \pm 0.5 \mathrm{km s^{-1} Mpc^{-1}}$) \cite{Planck2018}, indicating that our modified gravity framework remains consistent with $\Lambda$CDM cosmology at late times. This concordance suggests that the non-linear coupling between curvature and matter in our $f(R, L_m)$ model does not significantly alter the standard background expansion history. The present-day equation of state parameter, $\omega_0 \simeq -0.50$, indicates a dark energy behavior close to that of a cosmological constant ($\omega = -1$), but with slight deviations toward the quintessence regime ($\omega > -1$). This departure reflects the dynamical character of dark energy introduced by the oscillatory term $b \sin[\log(1+z)]$. The best-fit value, $b \simeq 0.95$, suggests mild oscillations in $\omega(z)$ over cosmic time, allowing the equation of state to periodically cross the phantom divide ($\omega = -1$) in a smooth and non-singular manner. Such behavior has been proposed in the literature as a plausible explanation for small residual features observed in late-time expansion data \cite{Amits24,SamaddarS2025,AmitSin2025}. The strong consistency observed across all three dataset combinations underscores the robustness of the model. The limited variation in parameter estimates from Pantheon+, DES-SN5Y, and Union 3 demonstrates that the oscillatory $f(R, L_m)$ framework provides a reliable and observationally viable alternative to $\Lambda$CDM, while naturally allowing for potential departures in the dark energy sector.

In summary, these results indicate that the non-linear $f(R, L_m)$ model with an oscillatory equation of state can effectively reproduce the $\Lambda$CDM background expansion at present, while introducing rich dynamical behavior that may become relevant at intermediate redshifts.
\subsection{Information criteria-based evaluation of cosmological models}\label{sec4.5}
\hspace{0.5cm} To quantify the comparative effectiveness of our $f(R, L_m)$ model against the standard $\Lambda$CDM scenario, we adopt two established information-theoretic measures: the Akaike Information Criterion (AIC) and the Bayesian Information Criterion (BIC). These criteria account for both the model’s goodness of fit and its complexity, penalizing models that include redundant free parameters. The AIC, originally proposed by Akaike \cite{Akaike74} and widely utilized in cosmology \cite{Liddle04, Ness13}, is defined as
\begin{equation}\label{33}
\mathrm{AIC} = \chi^2_{\rm min} + 2l,
\end{equation}
where $\chi^2_{\rm min}$ corresponds to the minimum chi-squared of the best-fit model and $l$ denotes the total number of free parameters. The relative performance of two models can be quantified using the difference in their AIC values, defined as
\begin{equation}\label{34}
\Delta \mathrm{AIC} = \big| \mathrm{AIC}_{\rm model} - \mathrm{AIC}_{\Lambda \rm CDM} \big|.
\end{equation}

A value of $\Delta \mathrm{AIC} < 2$ indicates that the two models are statistically consistent, $2 \leq \Delta \mathrm{AIC} < 4$ corresponds to mild evidence against the model with the higher AIC, and $\Delta \mathrm{AIC} \geq 4$ represents strong evidence disfavoring it.

The BIC \cite{Sch78} incorporates the dataset size $N$, applying a stronger penalty to models with greater complexity:
\begin{equation}\label{35}
\mathrm{BIC} = \chi^2_{\rm min} + l \ln N,
\end{equation}
where $\chi^2_{\rm min}$ is the minimum chi-squared value for the best-fit model and $l$ is the number of free parameters. The relative comparison between models is quantified as
\begin{equation}\label{36}
\Delta \mathrm{BIC} = \big| \mathrm{BIC}_{\rm model} - \mathrm{BIC}_{\Lambda \rm CDM} \big|.
\end{equation}

A value of $\Delta \mathrm{BIC} < 2$ indicates negligible evidence against the model, $2 \leq \Delta \mathrm{BIC} < 6$ corresponds to moderate evidence, $6 \leq \Delta \mathrm{BIC} < 10$ signals strong evidence, and $\Delta \mathrm{BIC} \geq 10$ represents very strong evidence disfavoring the model with the larger BIC.

From the results presented in Table \ref{Tab:T1}, the small differences in AIC values, $\Delta \mathrm{AIC} < 2$, indicate that our $f(R,L_m)$ model yields a statistically comparable fit to the $\Lambda$CDM model across all three dataset combinations. This suggests that the oscillatory equation of state does not add unnecessary complexity while remaining consistent with observational data. On the other hand, the relatively large $\Delta \mathrm{BIC} \simeq 11$ across all datasets provides very strong evidence favoring the simpler $\Lambda$CDM model when accounting for model complexity, as BIC applies a more stringent penalty. This arises because our model includes additional parameters ($\omega_0$, $b$) beyond those in $\Lambda$CDM. Overall, while the AIC results support the $f(R, L_m)$ model as a viable alternative, the BIC analysis emphasizes that the simpler $\Lambda$CDM model remains preferred under stricter statistical criteria.
\section{Physical implications of the fitted model parameters}\label{sec5}
\begin{figure}[hbt!]
  \centering
  \includegraphics[scale=0.42]{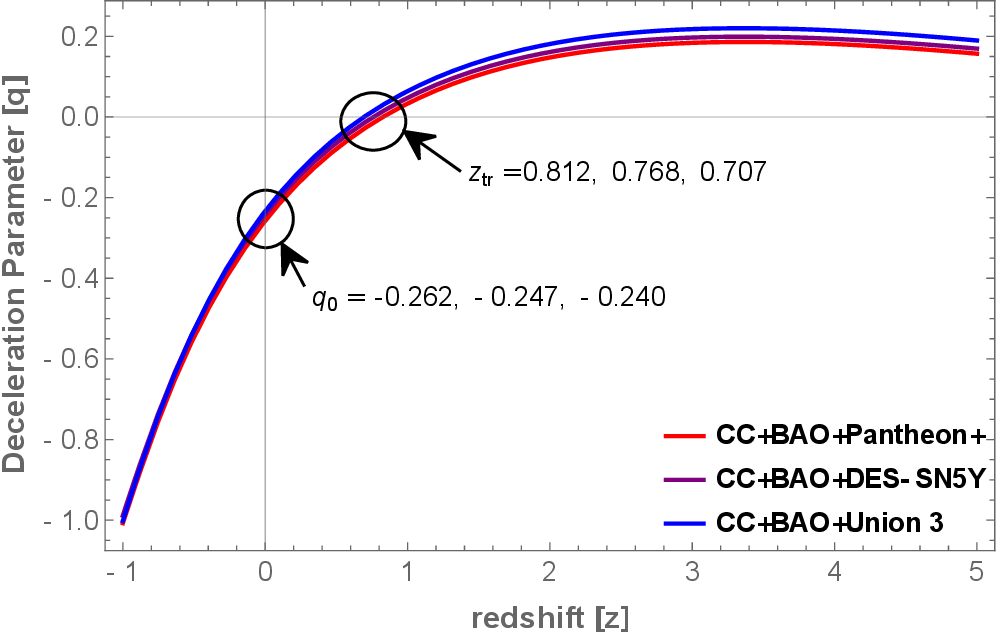}
  \caption{Variation of the deceleration parameter $q(z)$ with redshift $z$ for the $f(R, L_{m})$ model, using the CC+BAO+Pantheon+, CC+BAO+DES-SN5Y and CC+BAO+Union 3 datasets.
}\label{fig:f2}
\end{figure}

\hspace{0.5cm} The deceleration parameter $q$ provides a direct measure of the expansion dynamics of the Universe, distinguishing between accelerated $(q<0)$ and decelerated $(q>0)$ regimes. Defined via the second derivative of the scale factor $a(t)$, its sign reversal represents the transition from matter-driven deceleration to dark-energy-induced acceleration. By employing equation (\ref{20}) together with the MCMC-derived best-fit parameters, the deceleration parameter $q(z)$ can be expressed as a function of redshift. The resulting profiles of $q(z)$ versus $z$ for each dataset combination are illustrated in Figure \ref{fig:f2}.
\begin{equation}\label{37}
q(z)=-1+(1+\omega_{0})+b\sin(\log(1+z)).
\end{equation}

From Figure \ref{fig:f2}, we observe that at higher redshifts (early epochs), the curves begin with positive values of $q$, corresponding to a decelerated expansion phase. As the Universe evolves, $q(z)$ gradually decreases and crosses the transition point at $z_{tr}=0.812$, $0.768$ and $0.707$, for the CC+BAO+Pantheon+, CC+BAO+DES-SN5Y and CC+BAO+Union 3 datasets, respectively. Beyond these redshifts, $q$ becomes negative, indicating the onset of cosmic acceleration. At late times, all curves asymptotically approach $q \rightarrow -1$, representing the de-Sitter phase of expansion dominated by dark energy. The present-day values are obtained as $q_{0}=-0.262$, $-0.247$ and $-0.240$, for the CC+BAO+Pantheon+, CC+BAO+DES-SN5Y and CC+BAO+Union 3 datasets, respectively. These values are consistent with recent observational constraints, such as $q_{0} \approx -0.5 \pm 0.1$ from various cosmological probes, confirming that our $f(R,L_{m})$ model with the oscillatory EoS form remains in good agreement with current observational trends and successfully reproduces the late-time acceleration of the Universe.\\

\begin{figure}[hbt!]
  \centering
  \includegraphics[scale=0.42]{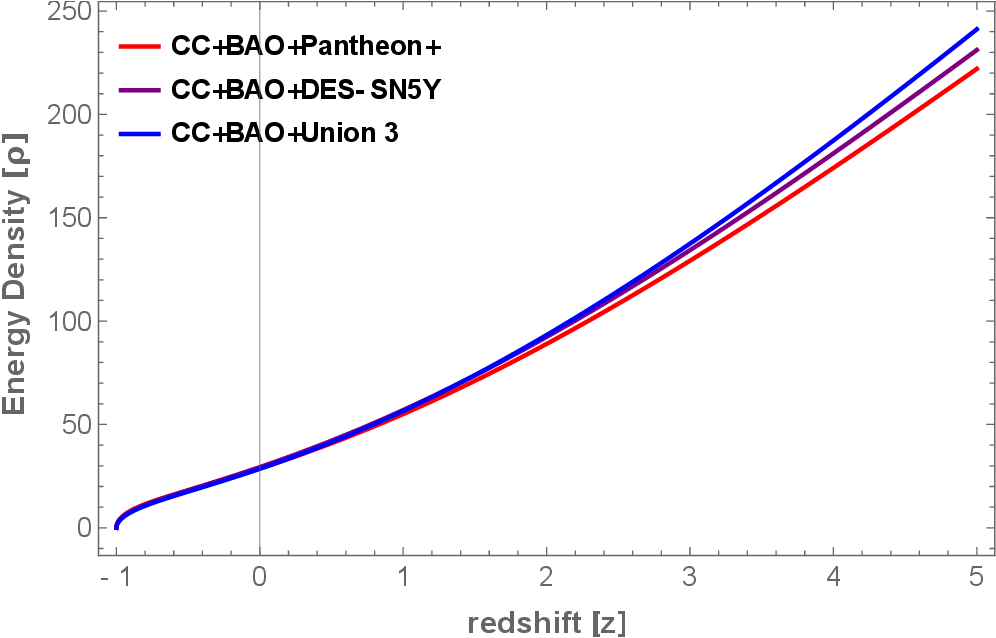}~~
  \includegraphics[scale=0.42]{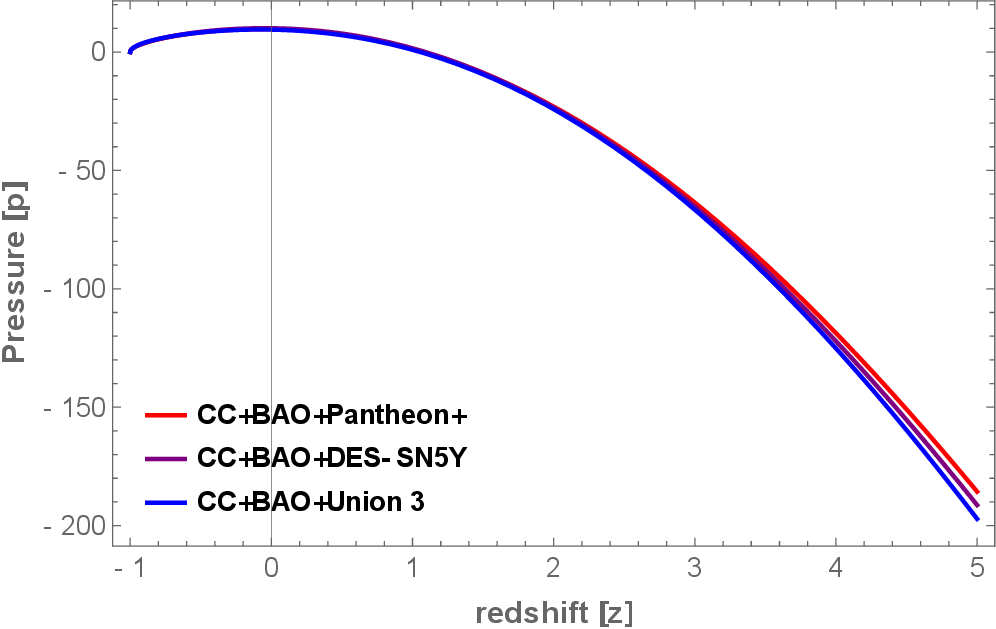}
  \caption{Behavior of the energy density $\rho(z)$ and pressure $p(z)$ versus redshift $z$ for the $f(R, L_{m})$ gravity model, reconstructed from the joint dataset combinations.}\label{fig:f3}
\end{figure}

Based on equations (\ref{17}) and (\ref{20}), the functional forms of the energy density $\rho(z)$ and pressure $p(z)$ are derived for the $f(R, L_{m})$ model featuring the oscillatory EoS. The variation of these quantities with redshift is presented in Figure \ref{fig:f3} for the three combined observational datasets considered in this analysis.
\begin{equation}\label{38}
\rho(z)=H_{0}(1+z)^{1+\omega_{0}}e^{-b\cos(\log(1+z))}\;,
\end{equation}
\begin{equation}\label{39}
p(z)=H_{0}(1+z)^{1+\omega_{0}}e^{-b\cos(\log(1+z))}\left[\omega_{0}+b\sin(\log(1+z))\right].
\end{equation}

From Figure \ref{fig:f3}, it can be observed that the energy density $\rho(z)$ remains strictly positive across the entire redshift range for all dataset combinations, confirming the physical consistency of the model. Conversely, the pressure $p(z)$ transitions to negative values at low redshifts, reflecting the late-time acceleration of the Universe typically associated with dark energy–dominated dynamics. The coexistence of positive energy density and negative pressure indicates that the model respects the essential energy conditions required for a viable cosmological framework. These results further emphasize that the $f(R, L_{m})$ model incorporating an oscillatory form of $\omega(z)$ offers a physically consistent and observationally supported description of the cosmic expansion history, aligning well with present-day cosmological data.\\

\begin{figure}[hbt!]
  \centering
  \includegraphics[scale=0.42]{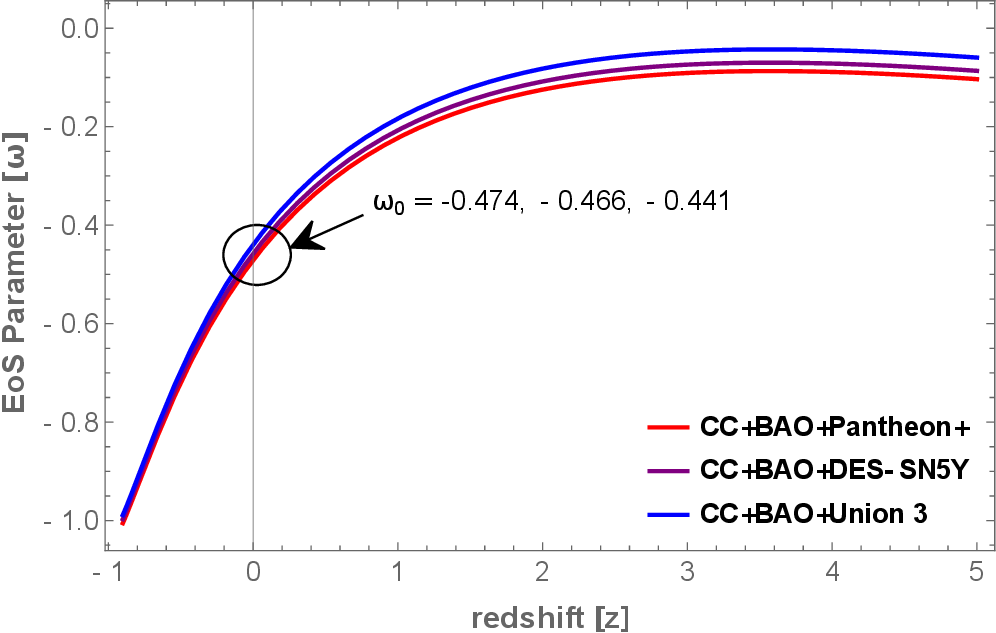}
  \caption{The redshift-dependent behavior of the $\omega(z)$ for the $f(R, L_{m})$ model, analyzed using the combined observational datasets: CC+BAO+Pantheon+, CC+BAO+DES-SN5Y and CC+BAO+Union 3 datasets.}\label{fig:f4}
\end{figure}

The oscillatory expression $\omega(z)=\omega_0+b \sin[\log(1+z)]]$ is used to study the nature of the EoS parameter $\omega(z)$. For each of the three combined observational datasets utilised in this work, the development of $\omega(z)$ is shown in Figure \ref{fig:f4}.

Figure \ref{fig:f4} shows that the current values of the EoS parameter are $\omega_0=-0.474$, $-0.466$ and $-0.441$ for the CC+BAO+Pantheon+, CC+BAO+DES-SN5Y and CC+BAO+Union 3 datasets, respectively. The close agreement with observational constraints confirms that the model provides an accurate description of the late-time dark energy dynamics. At early times (high redshifts), the EoS parameter $\omega(z)$ undergoes gentle oscillations due to the sinusoidal term. Toward the present epoch ($z \to 0$), all curves smoothly approach $\omega = -1$, reflecting a transition toward $\Lambda$CDM-like dynamics. This behaviour demonstrates that the oscillatory EoS inside the $f(R, L_{m})$ model can accurately describe the late-time accelerated expansion of the Universe and is still compatible with the observational evidence.\\

\begin{figure}[hbt!]
  \centering
  \includegraphics[scale=0.4]{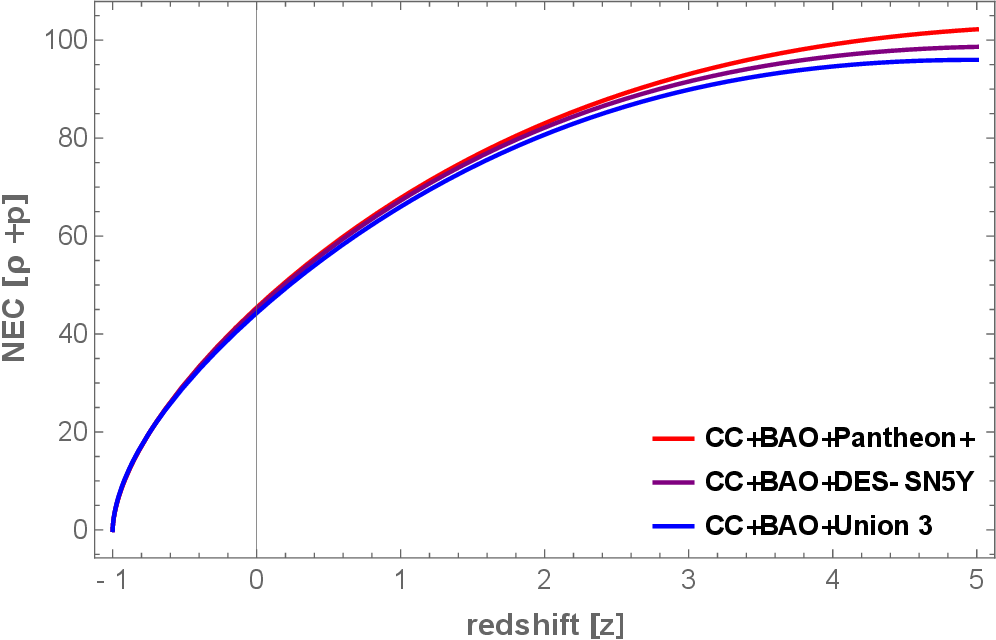}~~
  \includegraphics[scale=0.4]{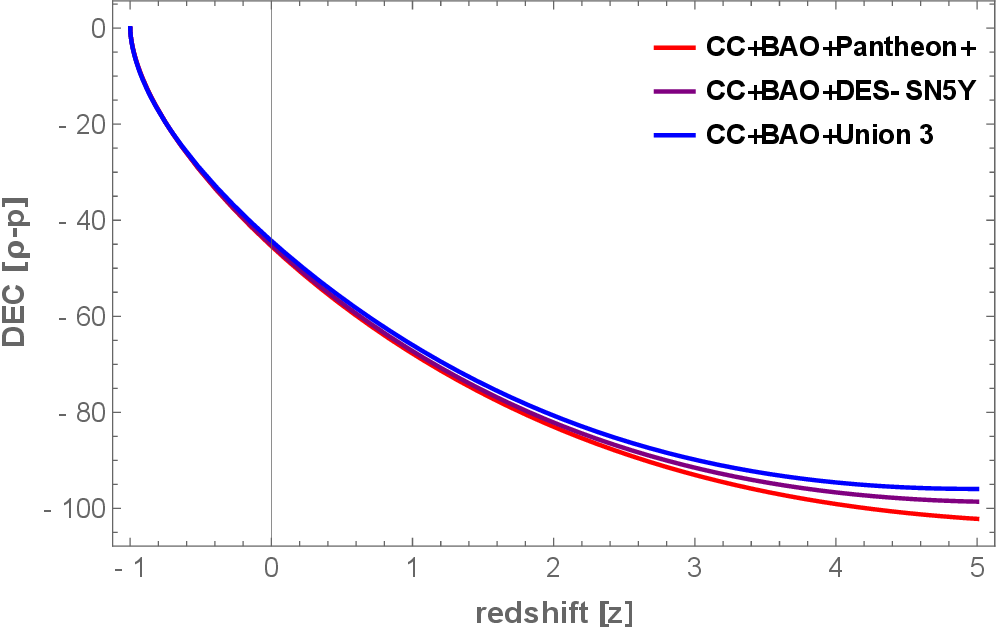}\\
  \vspace{0.2cm}
  \includegraphics[scale=0.4]{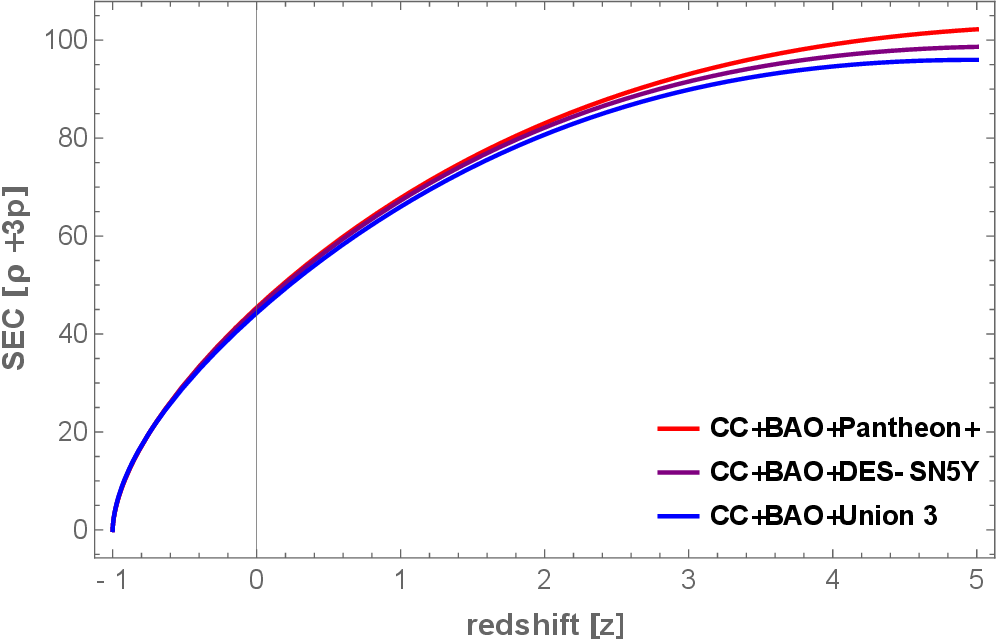}
  \caption{Evolution with redshift of the energy conditions (NEC, DEC and SEC) for the $f(R, L_{m})$ model, constrained using the CC+BAO+Pantheon+, CC+BAO+DES-SN5Y and CC+BAO+Union 3 datasets.}\label{fig:f5}
\end{figure}

To verify the physical plausibility of the $f(R, L_{m})$ model, we consider the standard energy conditions, which are crucial in general relativity and cosmology for ensuring viable energy–momentum configurations. In terms of the energy density $\rho$ and pressure $p$, these conditions are given by: Null Energy Condition (NEC): $\rho+p\geq0$, Dominant Energy Condition (DEC): $\rho-p\geq0$ and Strong Energy Condition (SEC): $\rho+3p\geq0$. The energy density and pressure are obtained from equations (\ref{37}) and (\ref{39}). Figure \ref{fig:f5} illustrates how these energy conditions evolve with redshift for the three combined observational datasets.

As $\rho+p\geq 0$ and $\rho\geq |p|$ remain positive throughout cosmic history, it is clear from Figure \ref{fig:f5} that the NEC and DEC meet the requirements for all three datasets. In contrast, the SEC, defined by $\rho+3p\geq 0$, is clearly violated. This violation of the SEC is physically significant, as it reflects the presence of negative effective pressure, a necessary feature for the observed cosmic acceleration. This behaviour further confirms that the oscillatory $f(R,L_{m})$ model permits a late-time accelerating universe while staying compatible with fundamental physical limitations. It is identical to the patterns found from the deceleration parameter and the EoS parameter.\\

To examine the geometric aspects of the $f(R,L_{m})$ gravity model and assess its distinction from competing dark energy formulations, we adopt the Statefinder approach developed by Sahni et al., which provides a higher-order characterization of cosmic dynamics extending past the usual Hubble and deceleration parameters. The Statefinder set $\{r, s\}$ is formulated as
\begin{equation}\label{40}
r = \frac{\dddot{a}}{a H^3}, \quad s = \frac{r - 1}{3(q - 1/2)},
\end{equation}
with $q$ being the deceleration parameter. This diagnostic serves as a framework-independent means of distinguishing dark energy models, where $\Lambda$CDM corresponds to the fixed point $(r, s) = (1, 0)$. Quintessence-like scenarios are associated with $r < 1$ and $s > 0$, while phantom-like scenarios correspond to $r > 1$ and $s < 0$.
\begin{figure}[hbt!]
  \centering
  \includegraphics[scale=0.42]{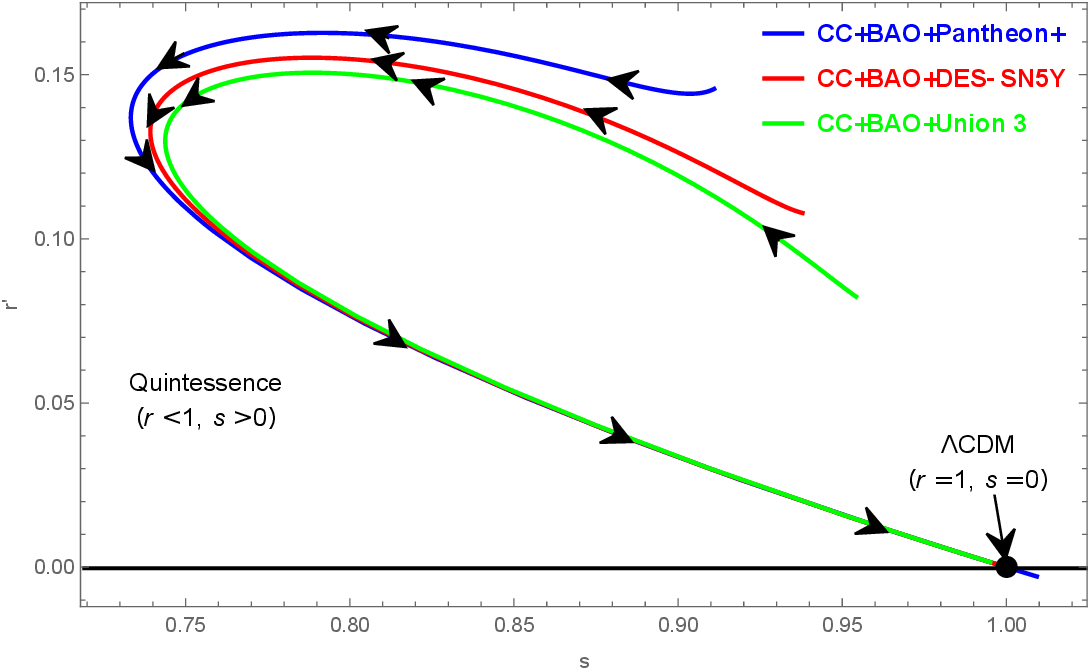}
  \caption{Evolution of the Statefinder parameters $\{r, s\}$ for the $f(R, L_{m})$ model.}\label{fig:f6}
\end{figure}

From Figure \ref{fig:f6}, it is observed that for all three datasets, the trajectories of the $\{r, s\}$ curves lie within the quintessence region, where $r<1$ and $s>0$ and eventually converge to the $\Lambda$CDM fixed point $(r=1, s=0)$ at late times. Evaluating the model at the current cosmic epoch gives the Statefinder values: $(r_0, s_0) = (0.779, 0.162), \; (0.756, 0.173), \; (0.807, 0.149)$, for the CC+BAO+Pantheon+, CC+BAO+DES-SN5Y and CC+BAO+Union 3 datasets, respectively. The consistent approach of all trajectories toward the $\Lambda$CDM point indicates that our model successfully reproduces the late-time cosmic acceleration while retaining a dynamical behavior characteristic of quintessence-like dark energy. This finding is in excellent agreement with the evolution of the deceleration parameter, EoS parameter and energy conditions, collectively confirming the viability and robustness of the $f(R, L_{m})$ model with an oscillatory EoS in describing the recent cosmic expansion history.
\begin{table}[h!]
\centering
\caption{Present cosmological quantities and corresponding transition redshifts for the analyzed datasets.}
\begin{tabular}{||p{2.9cm}|p{3.4cm}|p{3.8cm}|p{3.1cm}||}
\hline\hline
\hspace{0.5cm} Parameters & CC+BAO+Pantheon+ & CC+BAO+DES-SN5Y & CC+BAO+Union 3\\
\hline\hline
\hspace{1.2cm}$z_{tr}$ & \hspace{0.8cm}$0.812$ & \hspace{1.4cm}$0.768$ & \hspace{0.9cm}$0.707$\\[1.3pt]
\hline
\hspace{1.2cm}$q_{0}$ & \hspace{0.6cm}$-0.262$ & \hspace{1.2cm}$-0.247$ & \hspace{0.7cm}$-0.240$\\[1.3pt]
\hline
\hspace{1.2cm}$\omega_{0}$ & \hspace{0.6cm}$-0.474$ & \hspace{1.2cm}$-0.466$ & \hspace{0.7cm}$-0.441$ \\[1.3pt]
\hline 
\hspace{1.2cm}$r_{0}$ & \hspace{0.8cm}$0.779$ & \hspace{1.4cm}$0.756$ & \hspace{0.9cm}$0.807$\\[1.3pt]
\hline
\hspace{1.2cm}$s_{0}$ & \hspace{0.8cm}$0.162$ & \hspace{1.4cm}$0.173$ & \hspace{0.9cm}$0.149$\\[1.3pt]
\hline
\hspace{1.2cm}$t_{0}$ & \hspace{0.8cm}$13.30$ & \hspace{1.4cm}$13.39$ & \hspace{0.9cm}$13.23$\\
\hline\hline
\end{tabular}
\label{Tab:T2}
\end{table}

The Universe’s age, a key probe of its evolutionary timescale and structure formation, is computed in this framework by numerically integrating the reciprocal of the Hubble function over redshift:
\begin{equation}\label{41}
t_0 - t = \int_0^z \frac{dz'}{(1+z') H(z')},
\end{equation}
where $H(z)$ is the redshift-dependent Hubble parameter derived from our $f(R, L_{m})$ model. This approach allows a direct reconstruction of the cosmic timeline from the model’s expansion history.
\begin{figure}[hbt!]
  \centering
  \includegraphics[scale=0.42]{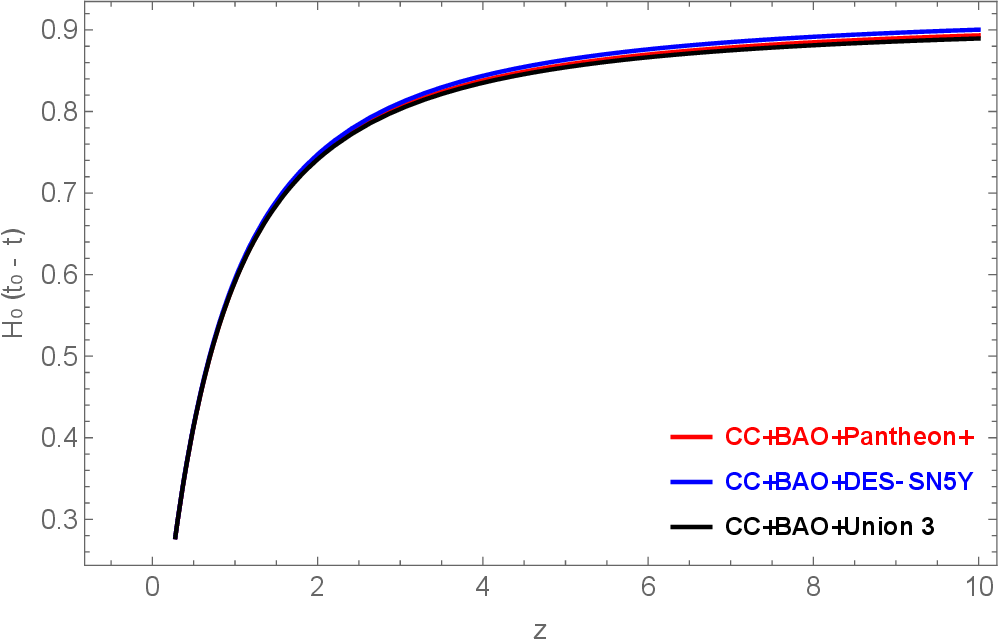}
  \caption{Comparison of the cosmic age evolution across redshift for the $f(R, L_{m})$ gravity model.}\label{fig:f7}
\end{figure}

From Figure \ref{fig:f7}, the estimated present-day ages of the Universe are $t_0 = 13.30, \; 13.39 \; \text{and} \; 13.23~\text{Gyr}$, for the CC+BAO+Pantheon+, CC+BAO+DES-SN5Y and CC+BAO+Union 3 datasets, respectively and these corresponding values are also summarized in Table \ref{Tab:T2} for reference.. These results show excellent agreement with independent cosmological probes, including the Planck 2018 measurements of the cosmic microwave background (CMB), which give $t_0 \approx 13.8$~Gyr. Furthermore, our findings are consistent with age estimates derived from the oldest globular clusters, white dwarf cooling sequences, and radioactive dating techniques \cite{Planck2018}. This agreement reinforces the reliability of the $f(R, L_{m})$ framework in reproducing realistic late-time cosmological behavior.
\section{Discussion and cosmological outlook}\label{sec6}
\hspace{0.5cm} In this study, we explored an oscillatory dark EoS parameterized as $\omega(z)=\omega_0+b\sin(\log(1+z))$, within the context of the non-linear $f(R,L_m)$ gravity framework characterized by $f(R, L_m) = \frac{R}{2} + L_m^2$. To tightly constrain the model parameters, a comprehensive MCMC analysis was carried out using a combination of the most recent Hubble cosmic chronometer (CC) dataset comprising 31 measurements, the DESI DR2 BAO data and three independent Type Ia supernova samples, Pantheon+, DES-SN5Y and Union 3, which ensure a statistically robust assessment of the model’s viability. The best-fit values obtained for these datasets, as presented in Table \ref{Tab:T1}, for the CC+BAO+Pantheon+, CC+BAO+DES-SN5Y, and CC+BAO+Union 3 datasets, respectively. The corresponding information criteria, $\Delta \mathrm{AIC} \approx 1.2$ and $\Delta \mathrm{BIC} \approx 11.4$, indicate that the model provides a statistically comparable fit to the standard $\Lambda$CDM scenario while allowing for richer late-time dynamics. The obtained Hubble constant values ($H_0 \simeq 67.2~\mathrm{km,s^{-1}Mpc^{-1}}$) exhibit excellent consistency with the Planck 2018 CMB results. Likewise, the present-day EoS parameter ($\omega_0 \approx -0.5$) agrees well with recent observational inferences from BAO, CMB, and SN~Ia data, implying a dark energy component that slightly departs from the $\Lambda$CDM prediction while remaining consistent with a quintessence-like evolution. The parameter value $b \approx 1$ indicates the existence of mild oscillatory features in $\omega(z)$, enabling the model to incorporate time-dependent variations in dark energy behavior without conflicting with current observational limits.

The evolution of the deceleration parameter $q(z)$ reveals a clear transition from an early decelerating epoch to a late-time accelerating phase, with transition redshifts $z_{\mathrm{tr}} = 0.812$, $0.768$ and $0.707$ for the three datasets, respectively. The present values $q_0=-0.262$, $-0.247$ and $-0.240$ are in good agreement with recent observational constraints, confirming that our $f(R, L_m)$ model with the oscillatory EoS effectively reproduces the observed late-time cosmic acceleration.

The energy density remains positive, while the pressure becomes negative at lower redshifts, signifying a physically consistent behavior leading to accelerated expansion. The EoS parameter evolves dynamically and asymptotically approaches $\omega=-1$ at late times, indicating a smooth transition to $\Lambda$CDM-like dynamics. The present EoS values $\omega_0 = -0.474$, $-0.466$ and $-0.441$ show excellent consistency with observational reconstructions from supernova and BAO data. Energy condition analyses reveal that the NEC and DEC are satisfied throughout cosmic evolution, while the SEC is violated, a characteristic indicator of dark energy–driven acceleration. This behavior aligns perfectly with the trends observed from the deceleration and EoS parameters, reinforcing the internal consistency of the model. The statefinder diagnostic further supports these results. All $(r, s)$ trajectories lie within the quintessence region ($r < 1$, $s > 0$) and converge toward the $\Lambda$CDM fixed point $(r= 1, s=0)$ at late times. The present-day statefinder values $(r_0, s_0) = (0.779, 0.162), (0.756, 0.173), (0.807, 0.149)$ confirm that the model effectively interpolates between quintessence-like dynamics and the cosmological constant regime as the Universe evolves.

Finally, the age of the Universe, estimated via the numerical integration of the inverse Hubble parameter, yields $t_0 = 13.30$, $13.39$ and $13.23$ Gyr for the three datasets (see Table \ref{Tab:T2}). These values exhibit excellent consistency with independent cosmological probes such as Planck 2018 and stellar age estimates from globular clusters and radioactive dating methods.

Overall, our results demonstrate that the non-linear $f(R, L_m)=\frac{R}{2}+L_m^2$ gravity model incorporating an oscillatory EoS form provides an observationally consistent, physically viable and dynamically rich description of the cosmic expansion history. The model reproduces the observed late-time acceleration, satisfies fundamental energy conditions and remains consistent with multiple cosmological datasets, presenting itself as a compelling alternative to the standard $\Lambda$CDM paradigm.


\begin{thebibliography}{99}
\bibitem{BP1} B.P. Abbott, Phys. Rev. Lett. {\bf116}, 061102 (2016).
\bibitem{BP2} B.P. Abbott, et al., Phys. Rev. Lett. {\bf116}, 241103 (2016).
\bibitem{BP3} B.P. Abbott, Phys. Rev. Lett. {\bf118}, 221101 (2017).
\bibitem{BP4} B.P. Abbott, Phys. Rev. Lett. {\bf119}, 141101 (2017).
\bibitem{BP5} B.P. Abbott, et al., Phys. Rev. Lett. {\bf119}, 161101 (2017).
\bibitem{SM2001} S.M. Carroll, Living Rev. Relat. {\bf4}, 1 (2001).
\bibitem{A1917} A. Einstein, Sitzungsber Preuss Akad Wiss Berlin (Math. Phys.) {\bf1917}, 142 (1917).
\bibitem{Kerner1982} R. Kerner, General Relativity and Gravitation {\bf14}, 453 (1982).
\bibitem{HA1970} H.A. Buchdahl, Monthly Notices of the Royal Astronomical Society {\bf150}, 1 (1970).
\bibitem{VF2004} V. Faraoni, Kluwer Academic, Dordrecht, 2004.
\bibitem{Bero06} O. Bertolami, J. Páramos, S. Turyshev, arXiv, arXiv:gr-qc/0602016.
\bibitem{Wang2012} J. Wang, K. Liao, Classical Quantum Gravity {\bf29}, 215016 (2012).
\bibitem{LV2022} L.V. Jaybhaye, et al., Phys. Lett. B {\bf831}, 137148 (2022).
\bibitem{LV2023} L.V. Jaybhaye, et al.,  Physics of the Dark Universe, {\bf40}, 101223 (2023).
\bibitem{singh2023} JK Singh, R Myrzakulov, H Balhara, New Astronomy, {\bf104}, 102070 (2023).
\bibitem{Maurya2023} DC Maurya, New Astronomy, {\bf100}, 101974 (2023).
\bibitem{Devi2024} YK Devi, SA Narawade, B Mishra, Physics of the Dark Universe, {\bf46}, 101640 (2024).
\bibitem{Tpad03} T. Padmanabhan, Physics Reports, {\bf380}, 235–320 (2003).
\bibitem{Chevallier2001} M. Chevallier and D. Polarski, International Journal of Modern Physics D, {\bf10}, 213–224 (2001).
\bibitem{Linder2003} E. V. Linder, Physical Review Letters, {\bf90}, 091301 (2003).
\bibitem{Huterer2001} T. Boehm and R. Brandenberger, Journal of Cosmology and Astroparticle Physics, {\bf2003}, 008 (2003).
\bibitem{Efstathiou1999} G. Efstathiou, Monthly Notices of the Royal Astronomical Society, {\bf310}, 842–850 (1999).
\bibitem{Barboza2008} E. M. Barboza and J. S. Alcaniz, Physics Letters B, {\bf666}, 415–419 (2008).
\bibitem{Jassal2005} H. K. Jassal, J. S. Bagla and T. Padmanabhan, Monthly Notices of the Royal Astronomical Society, 356, L11–L16 (2005).
\bibitem{Wetterich2004} C. Wetterich, Physics Letters B, {\bf594}, 17–22 (2004).
\bibitem{Gong2005} Y. Gong, Y.-Z. Zhang, Probing the curvature and dark energy, Physical Review D {\bf72}, (4) 043518 (2005).
\bibitem{Pan18} S. Pan et al., Phys. Rev. D {\bf98}, 063510 (2018).
\bibitem{Harko2010} T. Harko, F. S. N. Lobo, The European Physical Journal C, {\bf70}, 373–379 (2010).
\bibitem{Nmyr2023} N. Myrzakulov, et al., Eur. Phys. J. Plus 138 852 (2023).
\bibitem{Smyr2024} S. Myrzakulova, M. Koussour, N. Myrzakulov, Phys. Dark Univ. {\bf43}, 101399 (2024).
\bibitem{Lazkoz2010} R. Lazkoz, Phys. Lett. B, {\bf694}, 198 (2010).
\bibitem{Pace2012} F. Pace et al., MNRAS, {\bf422}, 1186 (2012).
\bibitem{Demianski2020} M. Demianski et al., Front. Astron. Space Sci., {\bf7}, 130 (2020).
\bibitem{Mackey13} D. F. Mackey et al., Publ. Astron. Soc. Pac., {\bf125}, 306 (2013).
\bibitem{M12} M. Moresco et al., J. Cosmol. Astropart. Phys. {\bf08}, 006(2012).
\bibitem{Sam24} A. Samaddar, S. S. Singh, Fortschritte der Physik, {\bf72} (6), 2400006 (2024).
\bibitem{ASamaddar25} A. Samaddar, S. Surendra Singh, High Energy Density Physics {\bf56}, 101216 (2025).
\bibitem{DESI25} DESI Collaboration: M. Abdul-Karim, et al., DESI DR2 Results II: Measurements of Baryon Acoustic Oscillations and Cosmological Constraints. ArXiv. https://arxiv.org/abs/2503.14738.
\bibitem{Brout22} D. Brout, D. Scolnic, B. Popovic, A. G. Riess, A. Carr, J. Zuntz, R. Kessler, T. M. Davis, S. Hinton, D. Jones, et al., The Astrophysical Journal {\bf938} (2) 110 (2022).
\bibitem{DES25} A. Adame, et al., Journal of Cosmology and Astroparticle Physics {\bf2025}, (02) 021 (2025).
\bibitem{Rubin25} D. Rubin, G. Aldering, M. Betoule, A. Fruchter, X. Huang, A. G. Kim, C. Lidman, E. Linder, S. Perlmutter, P. Ruiz Lapuente, et al., The Astrophysical Journal 986 (2) 231 (2025).
\bibitem{Planck2018} Planck Collaboration, Aghanim, N., et al., Astronomy $\&$ Astrophysics, {\bf641}, A6 (2020).
\bibitem{Amits24} Amit Samaddar, S. Surendra Singh, Journal of High Energy Astrophysics, {\bf48}, 100404 (2025). 
\bibitem{SamaddarS2025} A. Samaddar, S. S. Singh, Physics of the Dark Universe, {\bf50}, 102081 (2025).
\bibitem{AmitSin2025} A. Samaddar, S. Surendra Singh, Astrophys Space Sci {\bf370}, 92 (2025).
\bibitem{Akaike74} H Akaike, IEEE Trans. Autom. Control. {\bf19}, 716-723 (1974).
\bibitem{Liddle04} A. R. Liddle, Mon. Not. R. Astron. Soc., {\bf351}, L49–L53 (2004).
\bibitem{Ness13} S. Nesseris, J. Garc$\acute{i}$a-Bellido, JCAP 08, 036 (2013).
\bibitem{Sch78} G. Schwarz, Ann. Statist. {\bf6}, 461-464 (1978).
\bibitem{Planck2018} Planck Collaboration, Aghanim, N., et al., Astronomy $\&$ Astrophysics, {\bf641}, A6 (2020).
\end{thebibliography}
\end{document}